# Dimension- and Facet-Dependent Altermagnetic Biferroics and Ferromagnetic Biferroics and Triferroics in CrSb

*Long Zhang and Guoying Gao**

L. Zhang, G. Gao

School of Physics

Huazhong University of Science and Technology

Wuhan 430074, China

E-mail: guoying_gao@mail.hust.edu.cn

G. Gao

Wuhan National High Magnetic Field Center

Huazhong University of Science and Technology

Wuhan 430074, China

## Abstract

Altermagnets have recently garnered significant interest due to their vanishing net magnetic moment and non-relativistic momentum-dependent spin splitting. However, altermagnetic (AM) multiferroics especially triferroics remain scarce. We investigate the experimentally synthesized non-van der Waals CrSb as a model system to explore the effects of dimensionality and facet orientation on its ferroic properties. NiAs, MnP, wurtzite (WZ), zincblende (ZB), and rocksalt (RS) phases are considered. Using first-principles calculations, we predict the altermagnetism of CrSb in MnP phase which has comparable stability with experimental NiAs phase. Both NiAs- and MnP-phase (110) facets exhibit AM–ferroelastic (FC) biferroics, while the WZ-phase bulk and (001) facets host ferromagnetic (FM)–ferroelectric (FE) biferroics. Notably, the WZ-phase (110) facets are identified as FM–FE–FC triferroics, with moderate energy barriers of 0.129 and 0.363 eV/atom for FE and FC switching, respectively. Both FE and FC switching can reverse the AM spin splitting in antiferromagnetic (AFM) configurations while preserving the high spin polarization in FM states. The magnetic anisotropy is highly tunable, exhibiting either uniaxial or in-plane behavior depending on the phase, dimension, and facet. This work establishes a framework for designing AM multiferroics through polymorphic, dimensional, and facet engineering, offering promising avenues for multifunctional spintronic applications.

## 1. Introduction

The recent discovery of altermagnets[1] represents a significant advance in condensed matter physics, introducing a distinct class of collinear magnetic order beyond traditional magnetic classifications.[2-6] These materials exhibit compensated magnetic moments, break time-reversal symmetry, and display non-relativistic momentum-dependent spin splitting with *d*-/*g*-/*i*-wave symmetry.[7-9] By combining the advantages of ferromagnets (e.g., strong spin-dependent transport phenomena) and antiferromagnets (e.g., negligible stray fields and robustness against external perturbations),[10,11] altermagnets show great potential for high-density and low-power spintronic applications.

Multiferroicity as the coexistence of two or more ferroic orders such as ferroelectricity, ferromagnetism, or ferroelasticity, provides additional degrees of freedom for material control and enables next-generation low-power electronic and spintronic devices.[12,13] The breaking of time-reversal and spatial symmetries may allow for the emergence of magnetic and ferroelastic (FC)/ferroelectric (FE) orders, respectively. Recent interest has expanded to biferroic altermagnets, including $BiFeO_3$,[14] $MnPSe_3$,[15] and $CuWP_2S_6$,[16] etc. However, triferroic altermagnetic (AM) systems, where AM, FE, and FC orders are coupled, remain largely unexplored.[17]

Three-dimensional (3D) non-van der Waals (n-vdW) CrSb has recently attracted attention, primarily due to its NiAs-type phase, which demonstrates AM ordering with a high Néel temperature of ~700 K.[18] The zincblende (ZB) phase of CrSb, experimentally confirmed as ferromagnetic (FM) with a Curie temperature of ~400 K,[19] has also been predicted to exhibit half-metallic (HM) behavior.[20] Theoretical studies also suggested that the wurtzite (WZ)[21] and rocksalt (RS)[22] phases of CrSb are FM, with HM and metallic properties, respectively. Notably, the WZ, RS, and MnP-type polymorphs have been synthesized in related materials.[23-25] However, the MnP-type CrSb remains unexplored experimentally and theoretically. Altermagnets are identified by several key theoretical criteria:[3,26] (1) an even number of magnetic atoms per unit cell; (2) the absence of both $PT$ and $T\tau$ symmetries; (3) alternating spin-splitting signs across the Brillouin zone; and (4) opposite-spin sublattices connected by a real-space rotation transformation. The AM orders in NiAs,[27,28] WZ,[7] and MnP[29] phases fit these theoretical criteria, thereby establishing a direct connection to the emerging altermagnetism framework.

Modern micro- and nano-devices predominantly employ thin-film architectures, which are essential for high integration density, especially in the post-Moore era. Two-dimensional (2D) n-vdW materials, characterized by strong chemical bonding with substrates, exhibit superior interfacial thermal and electrical conduction[28,30] and can be efficiently identified from 3D material databases and synthesized via direct epitaxy.[31] The experimental use of few-layer n-vdW materials can achieve multiferroicity, such as single-layer $CuCrSe_2$,[32] atomically thin ε-$Fe_2O_3$,[33] wafer-scale one-unit-cell $Cr_2S_3$,[34] and Cu doped ZnO film.[35] As CrSb polymorphs are scaled down from 3D to 2D forms, a thorough investigation of their thickness- and orientation-dependent properties becomes crucial for future device applications.

In this work, we systematically investigate the dimension- and facet-dependent ferroic properties of CrSb polymorphs, building upon experimentally realized NiAs and ZB phases in CrSb bulk and thin films,[18,19,36,37] as well as structurally analogous WZ, RS, and MnP phases prepared in related systems.[23-25] Reducing dimensionality induces a transition from single-ferroic AM order in bulk NiAs/MnP phase to biferroic AM-FC behavior in the (110) facet. The WZ phase evolves from bulk FM–FE biferroics to FM–FE–FC triferroics in the (110) facet. Both FE and FC switching can reverse the AM spin splitting in antiferromagnetic (AFM) configurations, and preserve high spin polarization such as HM and unipolar magnetic semiconductor (UMS) or half-semiconductor (HSC) features in FM states. These findings highlight the potential of n-vdW multiferroic altermagnets and suggest promising avenues for advanced control and application of spin splitting in both theoretical and experimental studies.

## 2. Results and Discussion

### 2.1. Bulk ferroic properties

The crystal structures of bulk CrSb in the NiAs, WZ, ZB, RS, and MnP phases are displayed in Figure 1a-e. Their corresponding space groups and relaxed lattice parameters are provided in Table 1 and 2, respectively. The lattice parameters we calculated are close to those reported experimentally and theoretically.[18,22,38,39]

The magnetic ground states of these phases are of particular interest. The considered magnetic configurations are illustrated in Figure S1, and their energy comparisons with reference to the AFM state of MnP phase are summarized in Figure 1f. The AFM states of the NiAs and MnP phases are energetically favorable over their FM counterparts by 0.252 and 0.239 eV/Cr, respectively. In contrast, the FM states are more stable in the WZ, ZB, and RS phases, with energy advantages of 0.299, 0.334, and 0.107 eV/Cr, respectively. The NiAs and MnP phases exhibit nearly identical energies in their AFM states, differing by only 0.012 eV/Cr, indicating their comparable stability. The RS, ZB, and WZ phases are comparatively less stable, with energies exceeding that of the MnP phase by 0.472, 1.105, and 2.250 eV/Cr, respectively. The calculated formation energies of bulk CrSb in the NiAs, WZ, ZB, RS, and MnP phases are -4.65, -2.41, -3.56, -4.19, -4.66 eV/f.u., respectively. Notably, the NiAs and metastable ZB phases of CrSb have been experimentally synthesized,[18,19] while the WZ, RS, and MnP phases have been realized in analogous materials,[23-25] and thus the negative energies suggest that all polymorphs for CrSb are experimentally accessible. The stability of WZ-type, RS-type, and MnP-type CrSb are checked through phonon dispersion spectra utilizing the finite displacement approach within the PHONOPY package,[40] ab initio molecular dynamics (AIMD) simulations using the Nosé-Hoover thermostat[41] at 300 K. The phonon dispersion spectra (Figure S2a-c) show no imaginary modes, indicating the dynamical stability. The results of AIMD simulations (Figure S2d-f) exhibit that the total energies have slight fluctuations and the final snapshot of structures retain the intact structure without evident disruption, suggesting the thermal stability. These diverse magnetic states across stable polymorphs provide a rich material platform for exploring phase-dependent functionalities within a single compound.

The calculated ground-state magnetic moments of Cr atoms in CrSb are comparable in magnitude across the phases (Figure 2a), specifically, the values are ±2.699 $\mu_B$ (NiAs), 3.344 $\mu_B$ (WZ), 3.235 $\mu_B$ (ZB), 3.188 $\mu_B$ (RS), and ±2.784 $\mu_B$ (MnP), which also indicate their AFM or FM coupling. The magnetic moments of Sb atoms are 0.000 $\mu_B$ (NiAs), -0.312 $\mu_B$ (WZ), -0.290 $\mu_B$ (ZB), -0.179 $\mu_B$ (RS), and ±0.001 $\mu_B$ (MnP), these values are negligible compared to those of Cr atoms in CrSb.

Magnetic anisotropy was evaluated by comparing energies relative to the state with magnetization along the *z*-direction (Figure 2b). The NiAs and WZ phases exhibit perpendicular magnetic anisotropy (PMA), with energy increases by 0.30 and 0.09 meV/atom, respectively, when magnetization is aligned along the *x*-/*y*-axis. The ZB and RS phases are magnetically isotropic due to the lattice equivalence. The MnP phase has an easy axis along the *x*-direction, with energies along the *y*- and *z*-axes higher by 0.20 and 0.23 meV/atom, respectively. The PMA or uniaxial magnetic anisotropy is crucial for high-density magnetic storage and stable miniaturized spintronic devices.

To gain a deeper understanding of the magnetic anisotropy origins, atom- and orbital-resolved analyses were performed. Atom-resolved magnetic anisotropy energy (MAE) analysis (Figure 2c) indicates that the Sb atoms dominate the anisotropy contributions due to stronger spin–orbit coupling (SOC) resulting from their heavier atomic mass compared to Cr atoms. Further orbital-resolved analysis (Figure 2d–i) reveals that in the NiAs phase, hybridization between $p_z$ and $p_y$/$p_x$ orbitals favors magnetization along the *x*/*y* directions (Figure 2d and 2g). In the WZ phase, the $p_z$ and $p_y$/$p_x$ hybridization contributes positively to MAE along the *x*/*y* directions, while a negative contribution arises from the $p_y$ and $p_x$ hybridization (Figure 2e and 2h).

For the MnP phase (Figure 2f and 2i), the $p_y$ and $p_x$ hybridization is the primary factor stabilizing the easy axis along $x$-direction, with the $p_z$–$p_x$ and $p_y$–$p_x$ hybridizations providing positive and negative contributions, respectively. These moderate MAE values are advantageous for stabilizing long-range magnetic order and supporting spintronic applications.

The spin-resolved band structures and density of states (DOS) for bulk CrSb in both AFM and FM states are presented in Figures 3a-e and S3a-j. High-symmetry points are illustrated in Figure 1g-i and Table S1-S3. The AFM ground state of the NiAs phase exhibits AM spin splitting along the -M' –Γ' –M' and -D–P–D pathways (Figure 3a), which is consistent with the results previously reported.[18,27] In the FM state, the NiAs phase shows a near-HM character owing to more spin-up bands crossing the Fermi level (Figure S3b). The WZ phase exhibits AM spin splitting in the AFM state and intrinsic HM behavior in the FM ground state (Figures 3b and S3c,d). The ZB and RS phases display fully degenerate band structures in the AFM state, indicative of conventional antiferromagnetism, while their FM ground states exhibit HM character and moderate spin polarization, respectively (Figures 3c,d and S3e-h), which is in line with reported studies.[22,39] Interestingly, the MnP phase demonstrates AM splitting in its AFM ground state and low spin polarization in the FM state (Figures 3e and S3i,j), this is the first report on altermagnetism of CrSb in the MnP phase. The widespread presence of AM splitting and high spin polarization (including half-metallicity) across these CrSb polymorphs underscores their potential as versatile spin-polarized materials for magnetic memory.

In addition to the FM and AM properties, we further examined the FE character in these CrSb polymorphs, aiming to explore the potential for magnetoelectric coupling. A fundamental prerequisite for ferroelectricity is structural polarity, which is satisfied in the WZ phase due to the non-overlapping centers of positive and negative ions, generating a stable out-of-plane electrical polarization. In contrast, the centrosymmetric structures of other phases considered, where these charge centers coincide, preclude the emergence of ferroelectricity. This intrinsic FE polarization in the WZ phase opens a promising avenue for the electrical manipulation of both spin and charge carriers. The FE switching and energy barrier for bulk WZ CrSb are presented in Figure 4a–b. The calculated barrier is 0.365 eV/atom, which is comparable to those of conventional FE materials, such as $ReIrGe_2S_6$ and $GaFeO_3$,[42,43] suggesting experimental feasibility. To unravel the coupling between ferroelectricity and electronic structure, we calculated the spin-resolved band structures contributed by Cr atoms and spin-resolved DOS in both FM and AFM configurations for the polar ($P6_3mc$) and non-polar ($P6_3mmc$) states in Figures 4c–h and S4a-f. The FE polarization is along the out-of-plane direction. A critical question addressed is how the spin behavior evolves upon reversal of the electric polarization vector. In the FM configuration, the polar state exhibits HM behavior, whereas the non-polar state does not (Figures 4c–e and S4a-c). In the AFM configuration, the AM spin splitting is reversible: spin-up and spin-down bands swap during the FE transition (Figures 4f,h and S4d,f). This reversible spin-splitting phenomenon can be understood through a combined operation of parity reversal (associated with the FE switching) and time-reversal, accompanied by a specific fractional lattice translation. The intermediate mesophase exhibits conventional AFM character with fully degenerate spin bands (Figures 4g and S4e). The direct correlation between the direction of the electric dipole and the resultant AM spin splitting is pivotal, as it provides a clear and practical mechanism for external control. These results collectively demonstrate that FE switching in bulk WZ CrSb serves as an effective knob for manipulating spin states, thereby offering a robust pathway for controlling AM properties and tailoring charge transport performance in future spintronic devices.

### 2.2. Ferroic properties in the (001) plane

Given the presence of AM spin splitting in the bulk NiAs, WZ, and MnP phases, we further investigated the 2D ferroic properties of these systems. The atomic structures of the n-vdW (001)-oriented facets are shown in Figure 5a–c. For each phase, three kinds of atomic-layer thicknesses were considered; their space groups are listed in Table 1. The minimum number of layers is adopted with one unit cell.

The relative stability of FM and AFM states exhibits a clear dependence on both the phase and the slab thickness. As shown in Figure 5d-f, in the NiAs-phase (001) facet, the FM state is less stable than the AFM state by 0.174–0.202 eV/Cr across all thicknesses. Conversely, the WZ-phase (001) facet exhibits a FM ground state, with energy differences ranging from 0.185 to 0.233 eV/Cr relative to the AFM configuration. For the MnP-phase (001) facet, the FM state is also less favorable, with energy increases of 0.157–0.199 eV/Cr compared to the AFM state.

The magnetic moments of Cr atoms (Figure 6a) indicate a ferrimagnetic (FiM) state in the NiAs-phase (001) facets across varying layer thicknesses. For clarity and consistency in comparison, we refer to this FiM state as AFM in the Figures. The net magnetic moments per unit cell for the NiAs-phase (001) facet with four, five, and six atomic layers are 0.979, 4.405, and 3.905 $\mu_B$, respectively. In the WZ-phase (001) facet, interfacial effects lead to inequivalent Cr atoms, with atomic magnetic moments from 2.719 to 4.176 $\mu_B$ (Figure 6b). In the MnP-phase (001) facet, systems with six, seven, and eight atomic layers exhibit zero net magnetic moment but non-uniform Cr magnetic moments (Figure 6c), identifying them as fully compensated FiM systems–a special AFM case.

Thickness-dependent magnetic anisotropy is observed in the (001) facets (Figure 5g–i). In the NiAs (001) facet, the easy magnetization axis is in-plane for four and five atomic layers but switches to out-of-plane for six layers. The MAE values are -0.29, -0.19, and 0.47 meV/atom for four, five, and six layers, respectively (Figure 5g). A similar trend occurs in the WZ (001) facet (Figure 5h), with MAE values of 0.05, -0.11, and -0.37 meV/atom for six, five, and four layers, respectively. For the MnP (001) facet (Figure 5i), the easy axis aligns along the *x*-direction for six and seven atomic layers, with MAE values of -0.03 and -0.26 meV/atom, respectively. The eight-layer configuration exhibits a weak PMA of 0.02 meV/atom between the *x*- and *z*-directions. This tunability of magnetic anisotropy, including the observed spin-reorientation transitions with varying thickness, provides a powerful degree of freedom for designing low-dimensional spintronic devices with tailored magnetic properties.

The spin-resolved band structures and DOSs also show significant variation in Figures 6d-f and S5a-i. As shown in Figures 6d and S5a,d,g, the four-layer NiAs-phase (001) facet exhibits low spin polarization, while the five- and six-layer structures exhibit HM and near-HM behavior, respectively. Its three-layer structure shows completely degenerate bands (Figure S6) and conventional AFM behavior (not FiM) due to its $P$-6$m$2 space group and symmetric surface Cr atoms. The WZ-phase (001) facet displays robust HM characteristic irrespective of its thickness (Figures 6e and S5b,e,h), a highly desirable trait for applications requiring high spin-polarized current injection, such as magnetic tunnel junctions. In contrast, the MnP-phase (001) facet shows completely degenerate band structures across all atomic layer thicknesses (Figure 6f), indicating the absence of spin splitting in this orientation (Figure S5c,f,i).

The FE properties were specifically explored in the polar WZ-phase (001) facet. The FE switching and the corresponding energy barrier for this representative slab are illustrated in Figure 7a and 7b. Employing a thin four-atomic-layer structure as an instance, the polar structure has a space group of $P3m1$, while the non-polar structure belongs to $P\text{-}3m1$. A notable dimensional effect is observed: the energy barrier for FE switching decreases significantly from the bulk value of 0.365 eV/atom to a lower value of 0.129 eV/atom in this 2D limit. This suppression of the energy barrier can be attributed to the weakened interlayer restoring forces and enhanced structural flexibility in the ultrathin regime, which facilitates the atomic displacements required for polarization reversal. The intermediate mesophase along the switching path is located 0.108 eV/atom above the ground state. These reduced barriers signify not only a thermodynamically more favorable switching process but also suggest the potential for low-power and fast-switching device operation in thin-film architectures. This suggests that thin-film CrSb devices could operate with lower power consumption and higher switching speeds compared to their bulk counterparts. Examining the coupling between ferroelectricity and electronic structures, we find that in the FM ground state, both the polar and non-polar states exhibit HM behavior (Figures 7c-e and S7a-c) This indicates that the FE switching in the FM configuration can toggle between two distinct conductive states (e.g., for different logic functions) without degrading the perfect spin polarization, a highly desirable feature for spin-filtering applications. In the AFM state, however, the FE transition links to a more complex magnetic evolution: the polar states show FiM character with a net magnetic moment of 0.987 $\mu_B$ and reduced spin polarization, whereas the non-polar state exhibits conventional AFM order with degenerate spin channels (Figures 7f-h and S7d-f). This direct coupling, where the direction of electric polarization selectively stabilizes distinct magnetic orders (FM/FiM/AFM), offers a rich playground for multifunctional nano-devices. It enables the use of an electric field not merely as a binary (on/off) switch but also as a sophisticated control parameter to selectively activate disparate electronic and spin transport states within a single material platform.

**2.3. Ferroic properties in the (110) plane**

To elucidate the influence of facet orientation on ferroic properties, we also examined the (110)-oriented facets of CrSb in the NiAs, WZ and MnP phases with different thicknesses (Figure 8a–c). The magnetic ground state preferences persist along this orientation but with nuanced energy landscapes (Figure 8d–f). In the NiAs-phase (110) facet, the AFM state remains more stable across all thicknesses, with energy differences relative to the FM state ranging from -0.199 to -0.444 eV/Cr (Figure 8d). In contrast, the WZ-phase (110) facet consistently favors the FM state, which is lower in energy by 0.234–0.256 eV/Cr (Figure 8e). The MnP-phase (110) facet shows small energy differences between FM and AFM states, with the AFM configuration slightly more stable by 0.085–0.112 eV/Cr (Figure 8f).

Analysis of the Cr magnetic moments provides further insight into the magnetic order (Figures 9a–c). All NiAs-phase (110) configurations exhibit zero net magnetic moment (Figure 9a), distinguishing them from the FiM behavior observed in the (001) orientation (Figure 6a). The one- and two-layer structures are AFM, while the three-layer system forms a fully compensated FiM state due to inequivalent Cr magnetic moments. The WZ-phase (110) facet remains FM across all thicknesses, with Cr moments varying between 3.290 and 3.654 $\mu_B$ (Figure 9b). The MnP-phase

(110) facet shows zero net moment for all thicknesses, indicating fully compensated FiM order in the four- and six-layer configurations and AFM order in the five-layer system (Figure 9c).

Magnetic anisotropy is highly tunable with thickness and phase (Figure 8g–i). In the NiAs-phase (110) facet (Figure 8g), the easy axis lies along the *x*-direction in the one-atomic-layer case, and along the *y*-direction for the two- and three-atomic-layer configurations. The energy differences relative to the *z*-direction are -0.51 meV/atom (one layer), -0.24 meV/atom (two layers) and -0.04 meV/atom (three layers). The WZ-phase (110) facet exhibits PMA for all thicknesses, with MAE values of 2.06, 0.39, and 0.37 meV/atom for one, two, and three atomic layers (Figure 8h), which are much higher than the bulk value of 0.09 meV/atom (Figure 2b). In the MnP-type (110) facet, the easy axis aligns along the *y*-direction across all thicknesses (Figure 8i). The energy advantage over the *z*-direction is 0.28, 0.44, and 0.08 meV/atom for four, five, and six layers, respectively. For the *x*-direction, the energy is lower than the *z*-direction by 0.18 meV/atom in the four-layer structure, but higher by 0.70 and 0.41 meV/atom in the five- and six-layer systems, respectively. This strong uniaxial magnetic anisotropy is desirable for stabilizing magnetic order in ultra-thin films and promising for 2D magnetic memory devices with high thermal stability.

We further calculate the electronic band structures and DOSs of the (110) facts with respective magnetic ground states for the three phases (Figures 9d–f and S8a-i). The NiAs-phase (110) facets with one to three atomic layers exhibit AM spin splitting along the Γ–M path (Figure 9d). The WZ-phase (110) facets show near-perfect spin polarization across all thicknesses (Figures 9e and S8b,e,h). The one-atomic-layer structure exhibits characteristics of a UMS/HSC, with spin-up and spin-down gaps of 0.44 and 2.12 eV, respectively. The two- and three-atomic-layer structures exhibit HM behavior. The MnP-phase (110) facets also display AM spin splitting along the Γ–M path (Figures 9f and S8c,f,i). The *k*-resolved spin textures for CrSb (110) facets of NiAs and MnP phases in monolayer systems are calculated in Figure S9. Around the Fermi level, the spin-up and spin-down channels show distinction, the spin-splitting is evident near the Γ point, which is consistent with the spin-resolved band structures in the first sub-figure of Figures 9d and 9f. These results, especially the robust AM spin splitting in NiAs and MnP phase from bulk to (110) facets, highlight their potential for highly efficient spin-dependent applications in the 2D limit.

The one-atomic-layer WZ-phase (110) facet was selected as a representative system to probe its FE properties, owing to its structural polarity that inherently supports spontaneous electric polarization. As depicted in Figure 10a, the FE switching pathway involves a transition between a polar state (space group $Pmc2_1$) and a non-polar state (space group *Pmma*). The calculated energy barrier for this FE transition is 0.129 eV/atom (Figure 10b), which is comparable to that of (001) facet and significantly lower than the bulk counterpart. This relatively low energy barrier implies that the FE switching in the WZ-phase (110) facet can be easily induced under external stimuli, which is crucial for potential FE applications. Further deepening the implications, the electronic structure analysis is uncovered. In the FM ground configuration, the polar states exhibit high spin polarization with a small gap, whereas the intermediate non-polar state shows metallic behavior with low spin polarization (Figures 10c-e and S10a-c). This striking difference in electronic and magnetic properties between the polar and non-polar states indicates that FE switching can modulate both electrical conductivity and spin polarization. The polar states of AFM configurations exhibit reversible AM spin splitting. This implies that the spin ordering can be adjusted through FE switching. Meanwhile, the non-polar state behaves as a conventional antiferromagnet (Figures 10f-h and S10d-f). These findings collectively demonstrate the direct magnetoelectric coupling and the possibility of controlling spin order via FE switching in the WZ-phase (110) facet. However, the (110) facets of NiAs and MnP phases display no FE behavior. This can be attributed to their lacking of the requisite structural asymmetry, and their overlapping centers of positive and negative ions in both phases. The capability to electrically toggle between metallic and semiconducting states with different spin polarizations could pave the way for the development of advanced reconfigurable electronics and non-volatile memory devices, where precise control over electronic and magnetic properties is essential for improved performance and functionality.

Ferroelasticity refers to the ability of a material to reversibly change its spontaneous strain in response to external stress. This phenomenon is typically accompanied by structural phase transitions, which involve the rearrangement of atoms within the crystal lattice. As illustrated in Figure 11a, the FC switching is examined for WZ-phase (110) facet with one-atomic-layer structure. The intermediate mesophase has a space group of *Pm*. The energy barrier for FC transition is 0.363 eV/atom (Figure 11b) and comparable to known ferroelastics, such as 0.320 eV/atom of $BP_5$,[44] indicating both thermal stability (> 0.03 eV) and experimental feasibility (< 0.6 eV).[45] The FC signal intensity is calculated as $\varepsilon = | b/a - 1 | \times 100$ %.[13,46] The WZ-phase (110) facet displays a reversible FC strain of 10.5 %, which is consistent with typical values reported in VOCl (15.1 %)[46] and ZnS (12.8 %).[47] This combination of a feasible FC switching barrier and significant strain response underscores its promise for mechano-memory and strain-engineered applications, where mechanical deformation can be effectively used to encode information or modulate material properties, offering avenues for data transport and storage and material functionality enhancement. In the FM ground state, the FC and intermediate states retain high spin polarization (Figures 11c-e and S11a-c). This suggests that the robustness of the magnetic properties against mechanical deformation, which can be used for maintaining the overall magnetic performance of the material in practical applications. The intermediate mesophase is metallic, while the polar states exhibit a small band gap. Under the AFM ordering, the FC states show AM spin splitting, while the intermediate state is FiM with a net moment of 1.328 $\mu_B$ and low spin polarization (Figures 11f-h and S11d-f), illustrating another avenue for strain-mediated control over magnetic order.

For a comprehensive comparison, the FC and magnetic properties for the (110) facets of NiAs and MnP phases are further analyzed in Figure S12a-h, S13a-h, S14a-f, and S15a-f, respectively. The intermediate mesophase of the NiAs phase has a space group of *Pm*, while that of the MnP phase belongs to *P*1. Both structures exhibit considerably high energy barriers (Figure S12b and S13b), attributed to stronger interatomic bonding, and enhanced structural stability. These high energy barriers imply that FC switching in these phases would require more extreme stimuli, potentially limiting their operational feasibility under ambient conditions. Despite the challenging switching kinetics, the strength of FC response for NiAs-phase (110) facet is 58.1 %, which is comparable to GaSb (67.0 %)[45] and $BP_5$ (41.4 %).[44] This indicates that the NiAs-phase (110) facet can be advantageous for applications requiring high signal strength and resistance to external interference. The MnP-phase (110) facet exhibits a FC reversible strain of 6.8 %, which is comparable to GeSe (6.6 %) and SnS (4.9 %).[48] In the ground AFM configuration of NiAs-phase (110) facet, the FC states maintain AM character, while the intermediate states are FiM with net magnetic moments of 1.830 $\mu_B$, and exhibit high spin polarization. This high spin polarization arises from a pronounced difference in the electronic states across the Fermi level between the two spin channels (Figures S12f-h and S14d-f). The MnP-phase FC states are also AM, with a FiM HM intermediate mesophase having a net moment of 0.222 $\mu_B$ (Figures S13f-h and S15d-f). In the FM configuration, the FC and intermediate states of the NiAs-phase (110) facet are metallic with low and high spin polarization, respectively (Figures S12c-e and S14a-c). The MnP-phase (110) facet exhibits metallic FC states with reversible spin-polarized character, while its intermediate state is HM (Figures S13c-e and S15a-c). These findings suggest that the FC switching can mechanically tune the transport performance and enable strain-mediated control of magnetic order.

The rich interplay of magnetic properties with FE and FC behaviors across different phases (WZ, NiAs, MnP), dimensions, and orientations reveals that CrSb polymorphs offer a versatile toolkit for controlling spin and charge degrees of freedom. The emergence of triferroic and biferroic characters in these systems highlights their potential for next-generation multifunctional applications in spintronics, where electric control of magnetism is desired; electronics, where reconfigurable conductivity is key; and straintronics, where mechanical deformation modulates device functionality.

## 3. Conclusion

To conclude, this work presents a comprehensive investigation of the ferroic properties in bulk and 2D CrSb with different phases (NiAs, WZ, ZB, RS, and MnP) in both (001) and (110) facts. The MnP phase exhibits nearly same stability to the experimental NiAs phase. In terms of magnetic properties, the NiAs phase possesses AM and PMA characters in the bulk, FiM HM behavior in the (001) facets, and AM and fully compensated FiM features in the (110) facets. The MnP-phase bulk and (110) facets show AM splitting, while its (001) facets display completely degenerate spin bands. The WZ phase shows a FM HM and PMA nature in the bulk and (001) and (110) facets in the ground state, and a UMS/HSC character in one-atomic-layer (110) facet, along with robust AM spin splitting in AFM configurations. Both ZB and RS phases exhibit FM ground states with magnetic isotropy, and the bulk ZB phase is HM. The magnetic anisotropy is highly tunable, ranging from uniaxial to isotropic depending on the phase, dimension, and facet.

The NiAs-, WZ-, and MnP-phase (110) facets exhibit FC characteristics, while their bulk and (001) facets do not. The WZ-phase bulk and (001) facet are FM-FE biferroics, while the WZ-phase (110) facet is FM–FE–FC triferroics, with moderate FE and FC barriers of 0.129 and 0.363 eV/atom, respectively. Figure 12a summarizes the dimension- and facet-dependent biferroic and triferroic behaviors. As illustrated in Figure 12b, the FE and FC switching can maintain HM or high spin polarization in the FM WZ-phase bulk, and (001) and (110) facets, and reversibly flip the AM spin splitting in the AFM configurations.

Our results provide a material design strategy leveraging polymorphism and dimensional confinement to tailor functional properties. This study opens exciting pathways toward developing memory technologies based on electronic, magnetic, and mechanical degrees of freedom. We anticipate that these findings will stimulate further experimental and theoretical exploration of CrSb-based nanostructures and accelerate their integration into advanced functional devices.

## 4. Experimental Section

All the first-principles calculations were carried out by the density functional theory (DFT) in the Vienna ab initio Simulation Package (VASP).[49] With the projected augmented wave (PAW) method, the Perdew-Burke-Ernzerhof (PBE) exchange-correlation functional of the generalized gradient approximation (GGA) was utilized,[50] which has been proven to be agreement with previous experimental and theoretical studies for CrSb.[38,51] The energy barriers were calculated using the climbing image nudged elastic band (CI-NEB) approach.[52] The convergences tolerances for force and energy were 0.001 eV Å$^{-1}$ and $10^{-6}$ eV, respectively. A 15 Å vacuum was employed to eliminate the interactions from the periodic structures in 2D phases. The Γ-centered Monkhorst-Pack grid was employed as follows: 10 × 10 × 8 for bulk NiAs and WZ phases; 8 × 8 × 8 for bulk ZB and RS phases; 8 × 11 × 6 for bulk MnP phase; 10 × 10 × 1 for NiAs- and WZ-phase (001) facets; 8 × 10 × 1 for MnP-phase (001) facets; 8 × 6 × 1 for NiAs- and WZ-phase (110) facets; and 6 × 6 × 1 for MnP-phase (110) facets.

**Supporting Information**

Supporting Information is available from the Wiley Online Library or from the author.

**Acknowledgements**

We gratefully acknowledge the insightful discussions of Prof. Menghao Wu from Huazhong University of Science and Technology and Asst. Prof. Zhuang Ma from Zhoukou Normal University on ferroelasticity and ferroelectricity. Guoying Gao acknowledges support from the National Natural Science Foundation of China (Grant No. 12174127).

**Conflict of Interest**

The authors declare no conflict of interest.

**Data Availability Statement**

The data that support the findings of this study are available from the corresponding author upon reasonable request.

## Figures and Tables

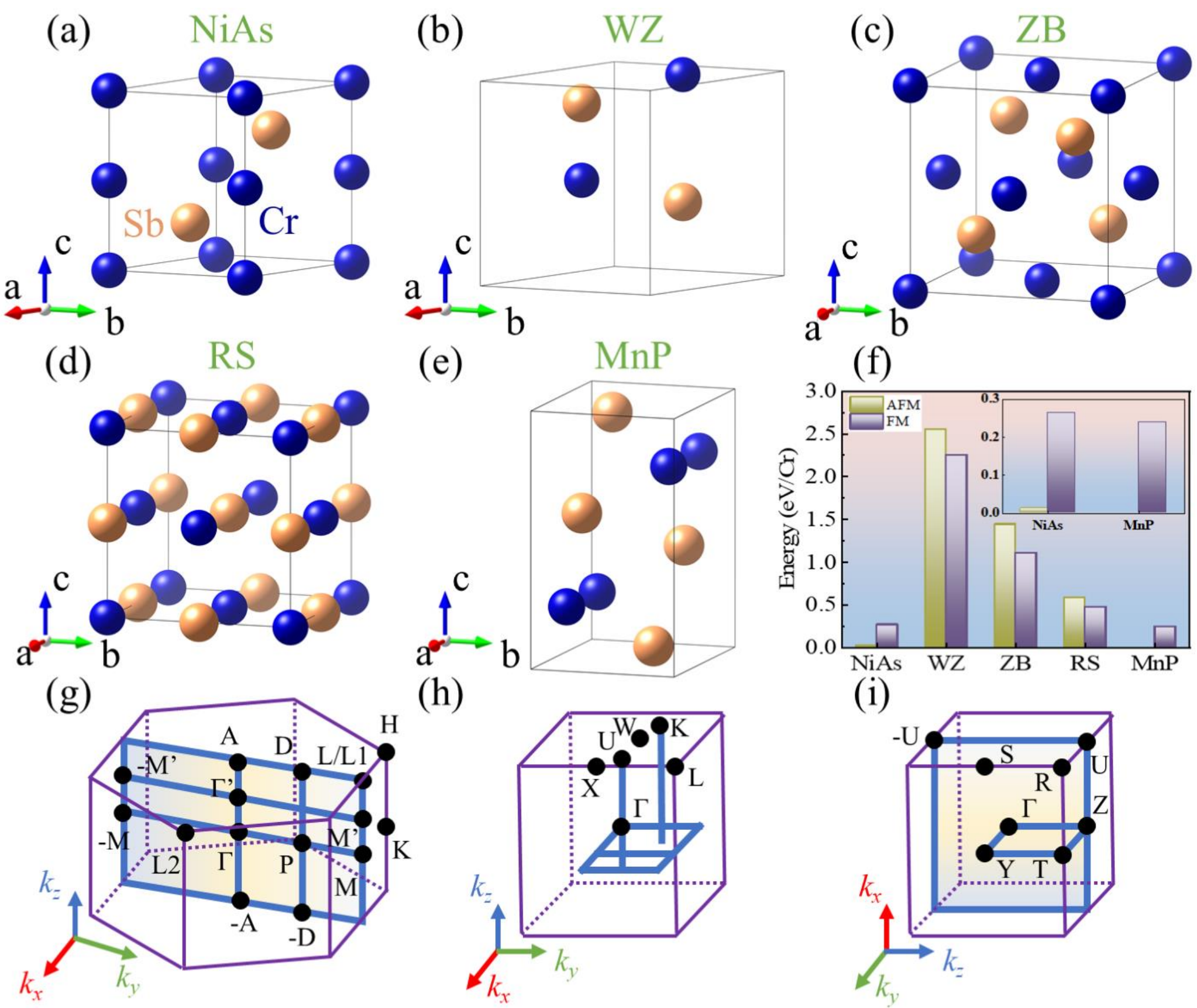

**Figure 1.** Crystal structures of bulk CrSb in the NiAs (a), wurtzite (WZ) (b), zincblende (ZB) (c), rocksalt (RS) (d), and MnP (e) phases. Relative energies of the antiferromagnetic (AFM) and ferromagnetic (FM) states, referenced to the AFM state of the MnP phase (f). Schematic diagram of high-symmetry points within Brillouin zone for NiAs and WZ (g), ZB and RS (h), and MnP (i) phases.

**Table 1.** Space groups of CrSb for bulk, (001) facet, and (110) facet configurations across different phases (NiAs, WZ, ZB, RS, and MnP). For the facets, the three entries correspond to the three kinds of atomic-layer thicknesses studied.

| Phase | Bulk | (001) facet | (110) facet |
|---|---|---|---|
| NiAs | $P6_3mmc$ | *P*3*m*1/*P*-3*m*1/*P*3*m*1 | *Pmma*/*Pbcm*/*Pmma* |
| WZ | $P6_3mc$ | *P*3*m*1/*P*3*m*1/*P*3*m*1 | $Pmc2_1/Pca2_1/Pmc2_1$ |
| ZB | $F\text{-}4_3m$ | - | - |
| RS | *Fm*-3*m* | - | - |
| MnP | *Pnma* | *Pmmn*/*Pmm*2/*Pmm*2 | *Pc*/*P*2*c*/*Pc* |

**Table 2.** Lattice parameters for bulk, (001) facet, and (110) facet configurations of CrSb with different phases. For the facets, the three entries correspond to the three kinds of atomic-layer thicknesses studied.

| Phase | Bulk (Å) | | | (001) facet (Å) | | (110) facet (Å) | |
|---|---|---|---|---|---|---|---|
| | *a* | *b* | *c* | *a* | *b* | *a* | *b* |
| NiAs | 4.147 | 4.147 | 5.265 | 4.146/4.188/4.162 | 4.146/4.188/4.162 | 4.448/5.132/5.309 | 7.032/7.290/7.176 |
| WZ | 5.248 | 5.248 | 6.071 | 4.386/4.425/4.421 | 4.386/4.425/4.421 | 6.708/6.941/6.957 | 7.411/7.181/7.213 |
| ZB | 6.122 | 6.122 | 6.122 | - | - | - | - |
| RS | 5.538 | 5.538 | 5.538 | - | - | - | - |
| MnP | 5.344 | 4.187 | 7.243 | 5.369/5.251/5.396 | 4.129/4.120/4.127 | 7.020/7.050/7.078 | 6.571/6.564/6.686 |

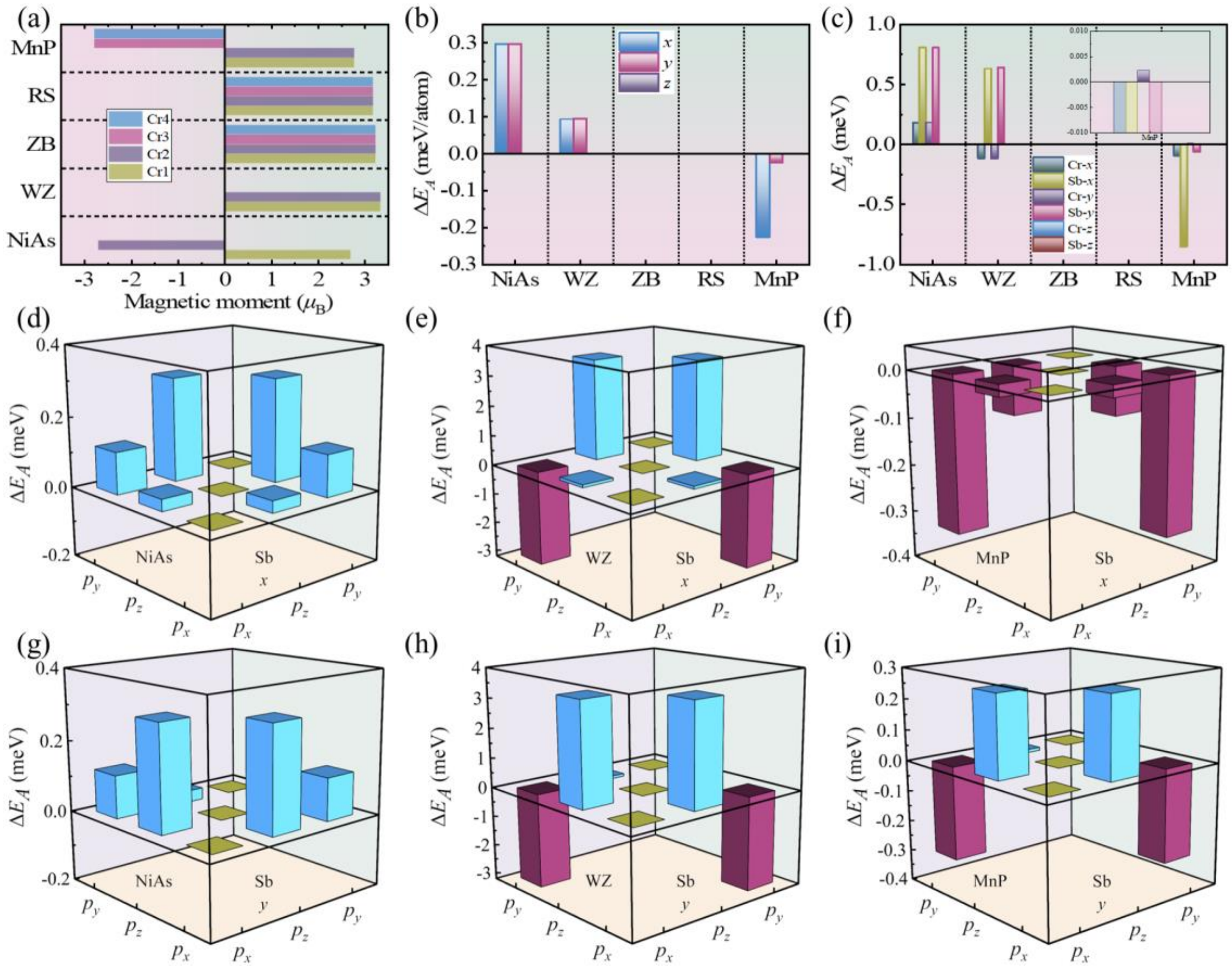

**Figure 2.** Atomic magnetic moments of Cr in CrSb with different phases in magnetic ground states (a). The total (b), atom-resolved (c), and Sb-orbital-resolved (d-i) magnetic anisotropy energy differences ($\Delta E_A$s), presented as the energy difference relative to the state with magnetization along *z*-direction.

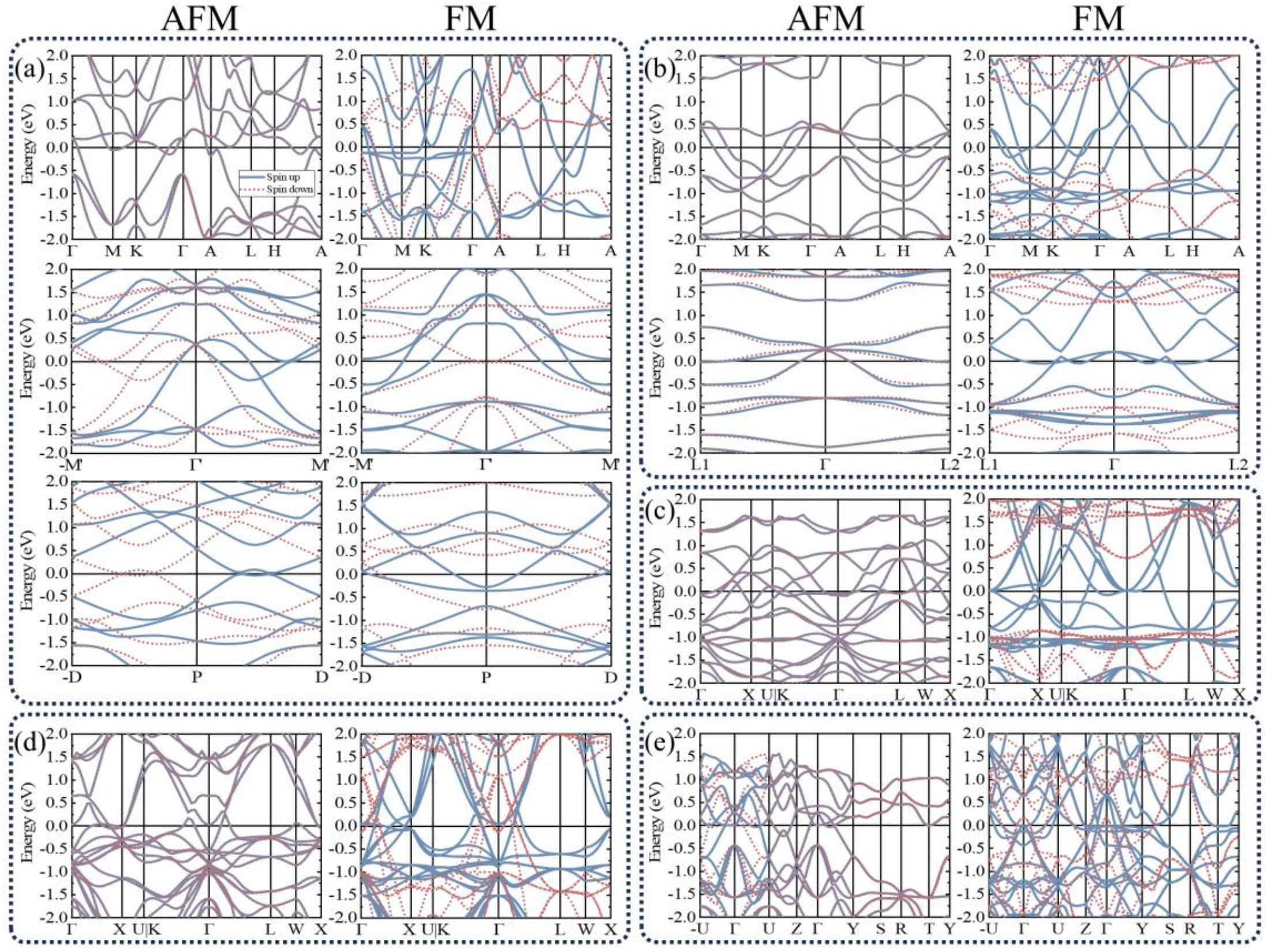

**Figure 3.** Spin-resolved band structures of bulk CrSb with the NiAs (a), WZ (b), ZB (c), RS (d), and MnP (e) phases.

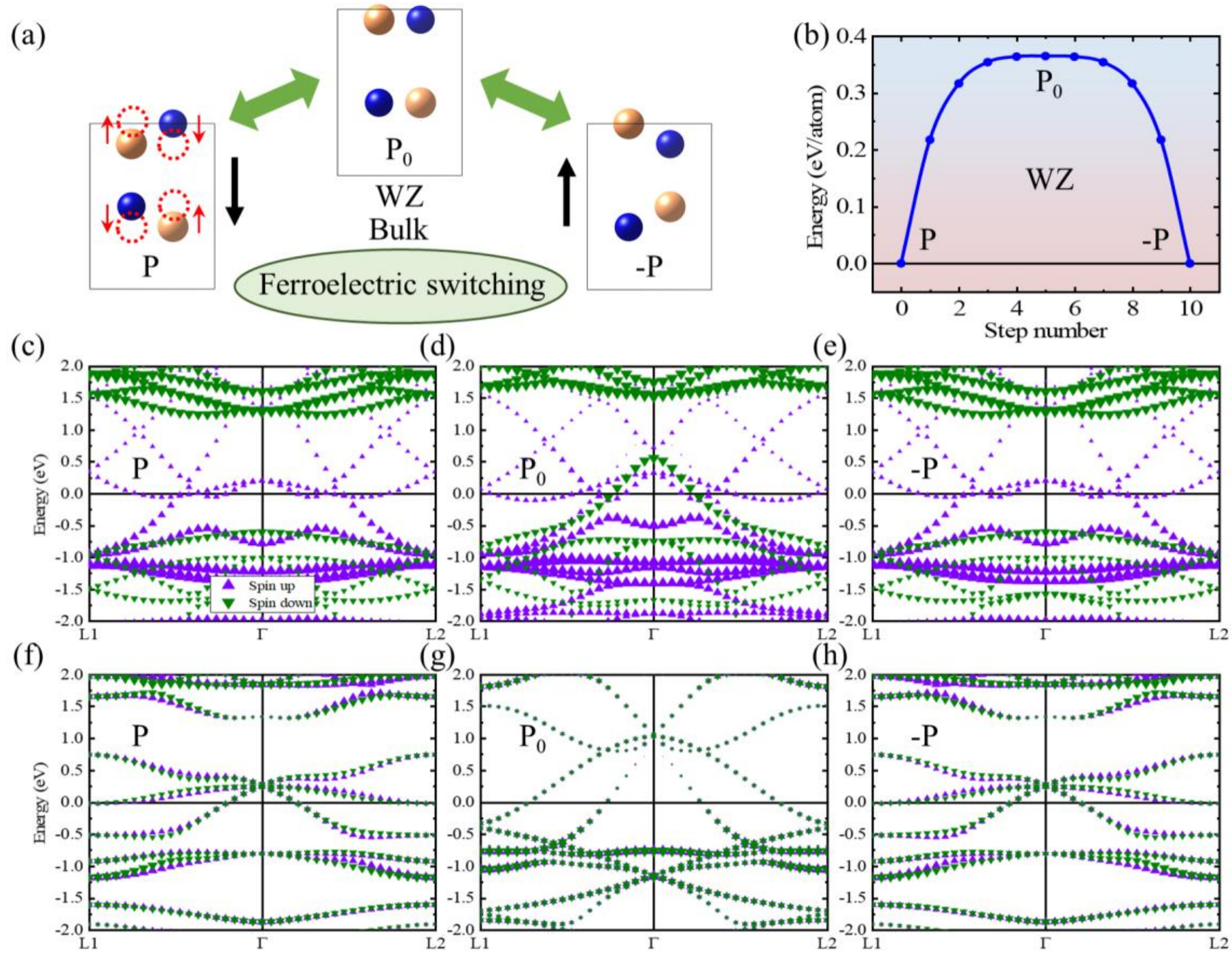

**Figure 4.** Ferroelectric (FE) switching with atomic displacement paths (a) and energy barrier for the FE transition (b) in bulk WZ-phase CrSb. Spin-resolved and Cr-contributed band structures for the FM (c-e) and AFM (f-h) configurations with the polar (P, -P) and non-polar ($P_0$) states.

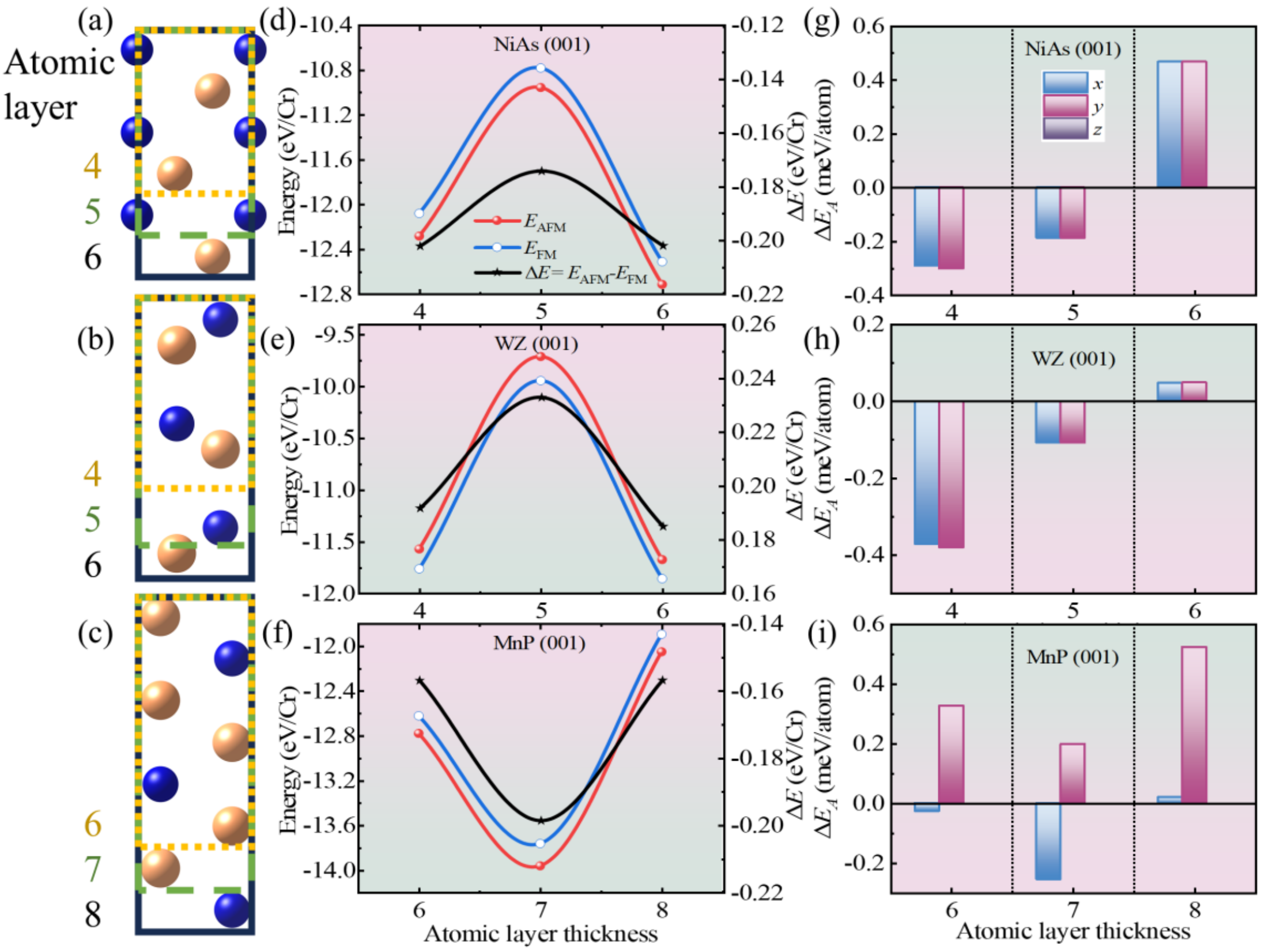

**Figure 5.** Crystal structures of 2D CrSb (001) facets in NiAs, WZ, and MnP phases with varying atomic-layer thicknesses (a-c), relative energies ($\Delta E$s) of AFM and FM states (d-f), and magnetic anisotropic energies ($\Delta E_A$s) with reference to the state with magnetization axes along $z$-direction (g-i).

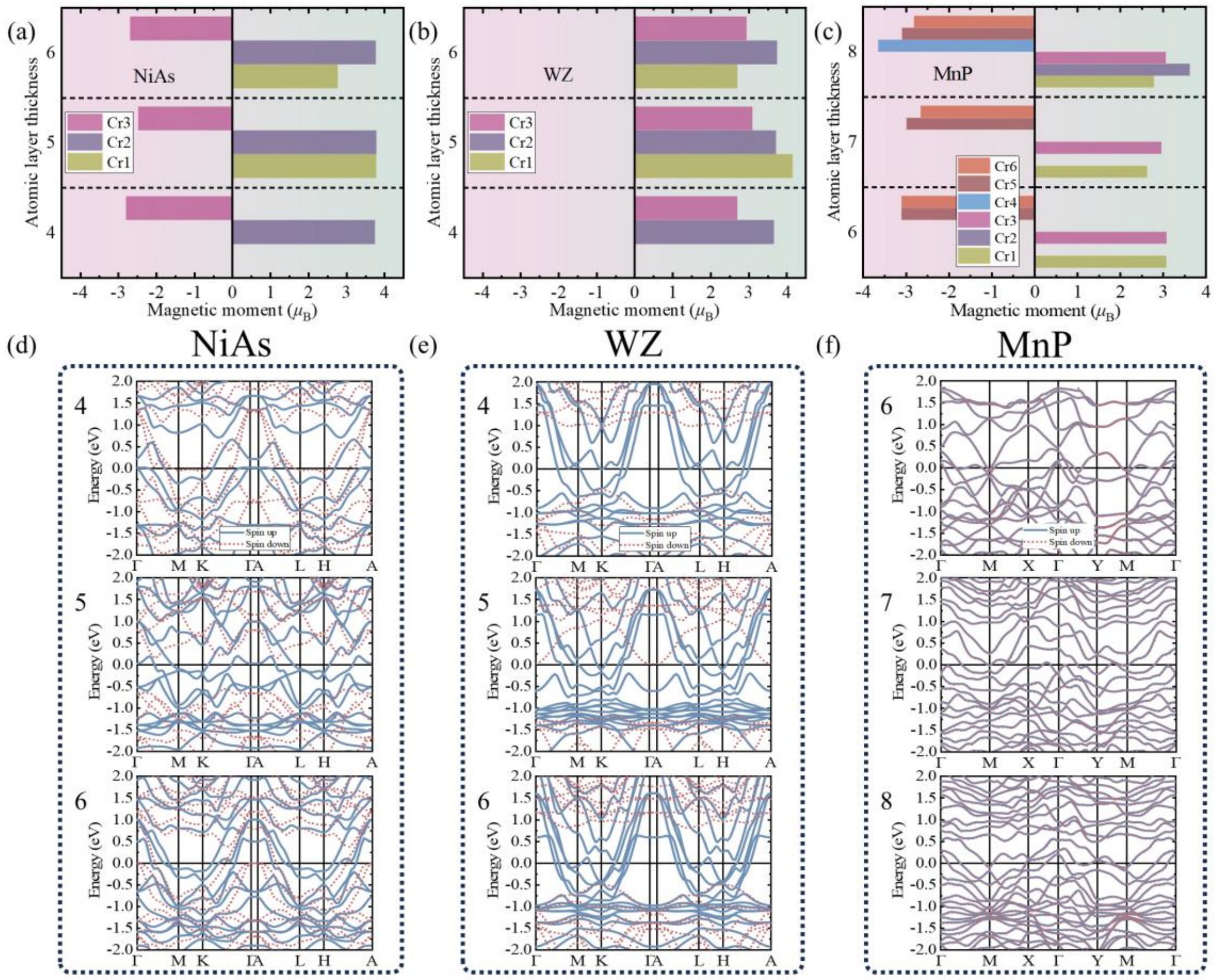

**Figure 6.** Atomic magnetic moments of Cr in CrSb (001) facet with varying atomic-layer thicknesses in the NiAs (a), WZ (b), and MnP (c) phases, and the corresponding spin-resolved band structures (d-f).

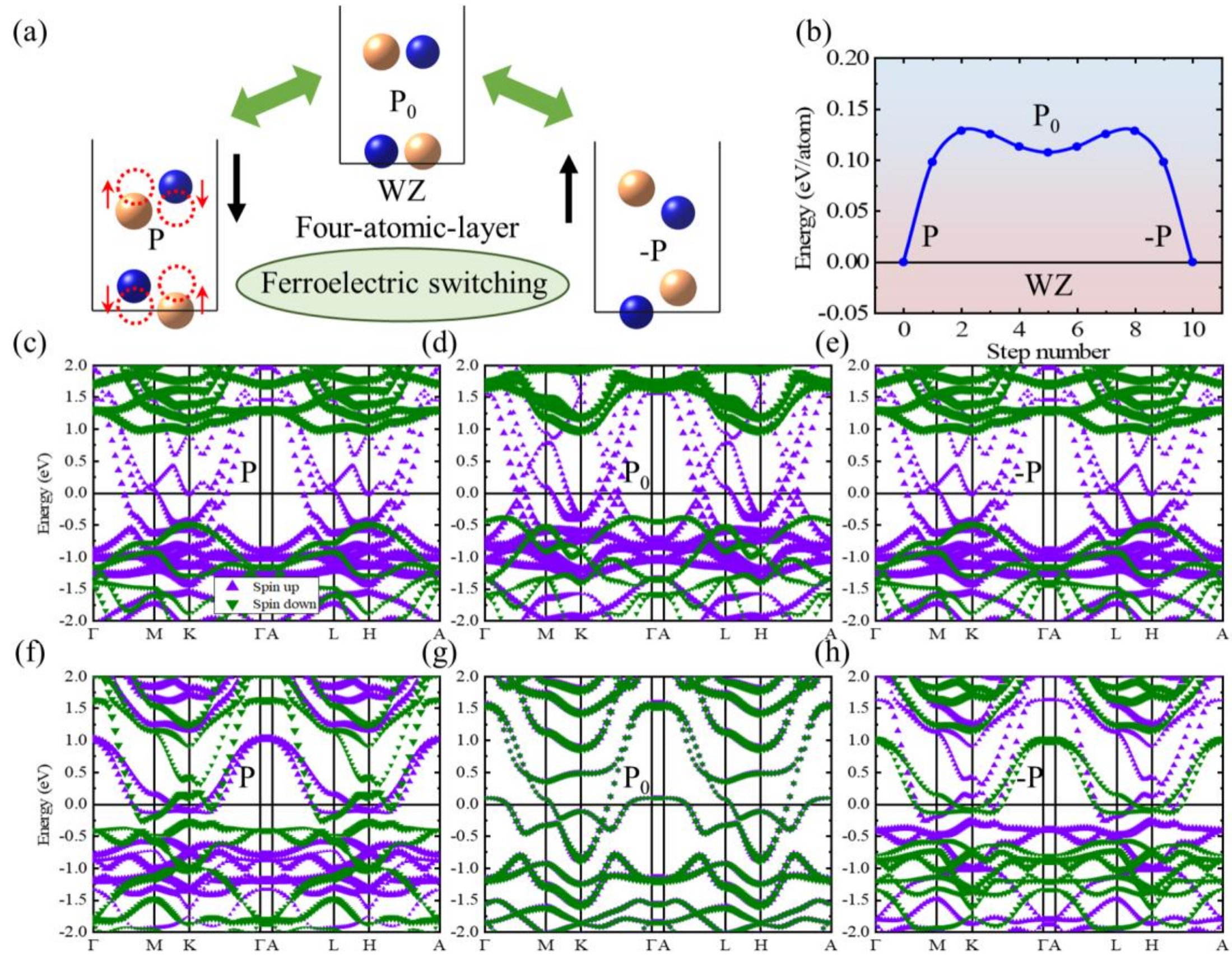

**Figure 7.** FE switching with atomic displacement paths (a) and energy barrier for the FE transition (b) in WZ-phase CrSb (001) facet with four atomic layers. Spin-resolved and Cr-contributed band structures for the FM (c-e) and AFM (f-h) configurations in the polar (P, -P) and non-polar ($P_0$) states.

.

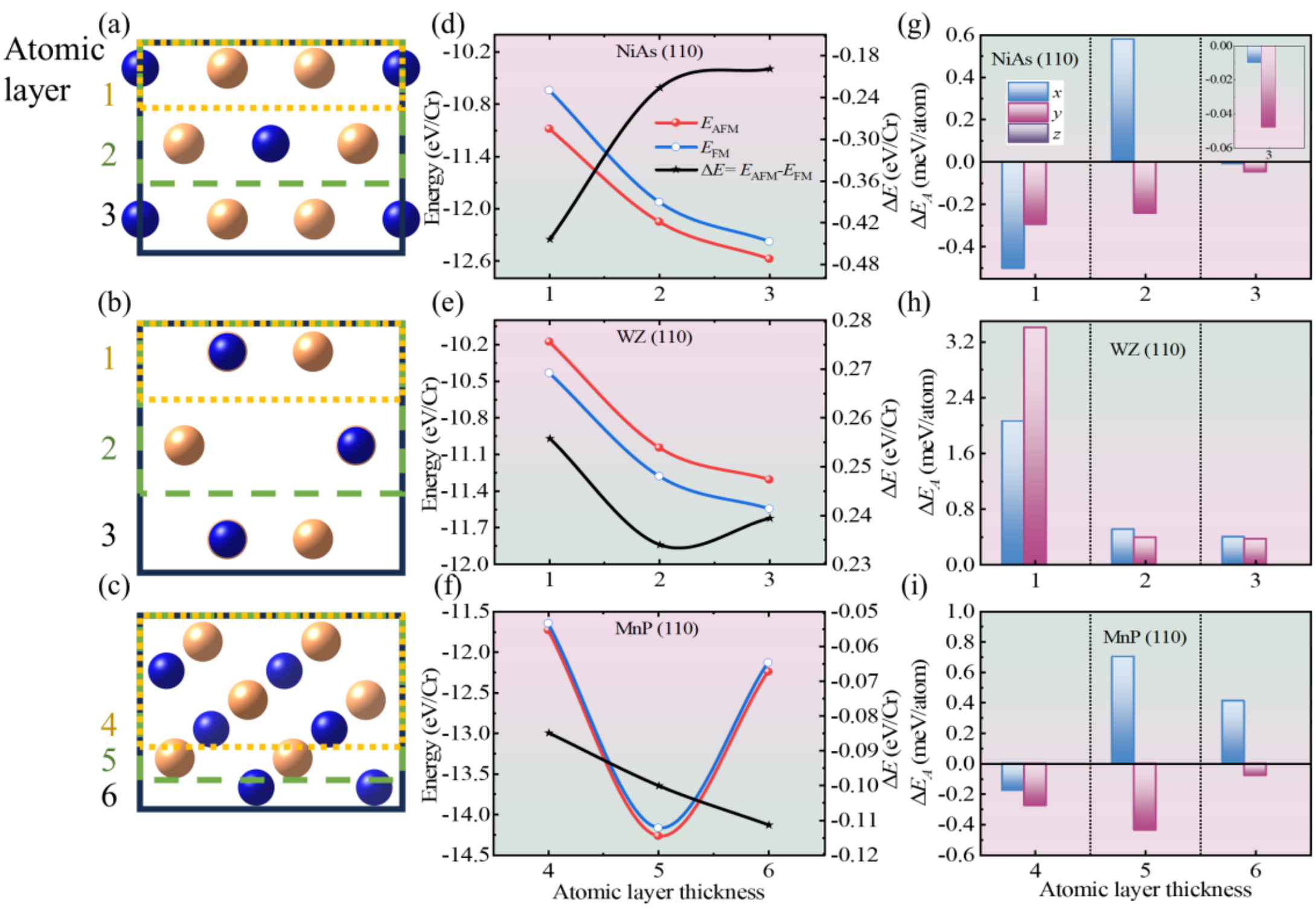

**Figure 8.** Crystal structures of 2D CrSb (110) facets with varying atomic-layer thicknesses (a-c), the relative energies of AFM and FM states (d-f), and magnetic anisotropic energies ($\Delta E_A$s) with reference to the state with magnetization axe along *z*-direction (g-i) for the NiAs, WZ, and MnP phases.

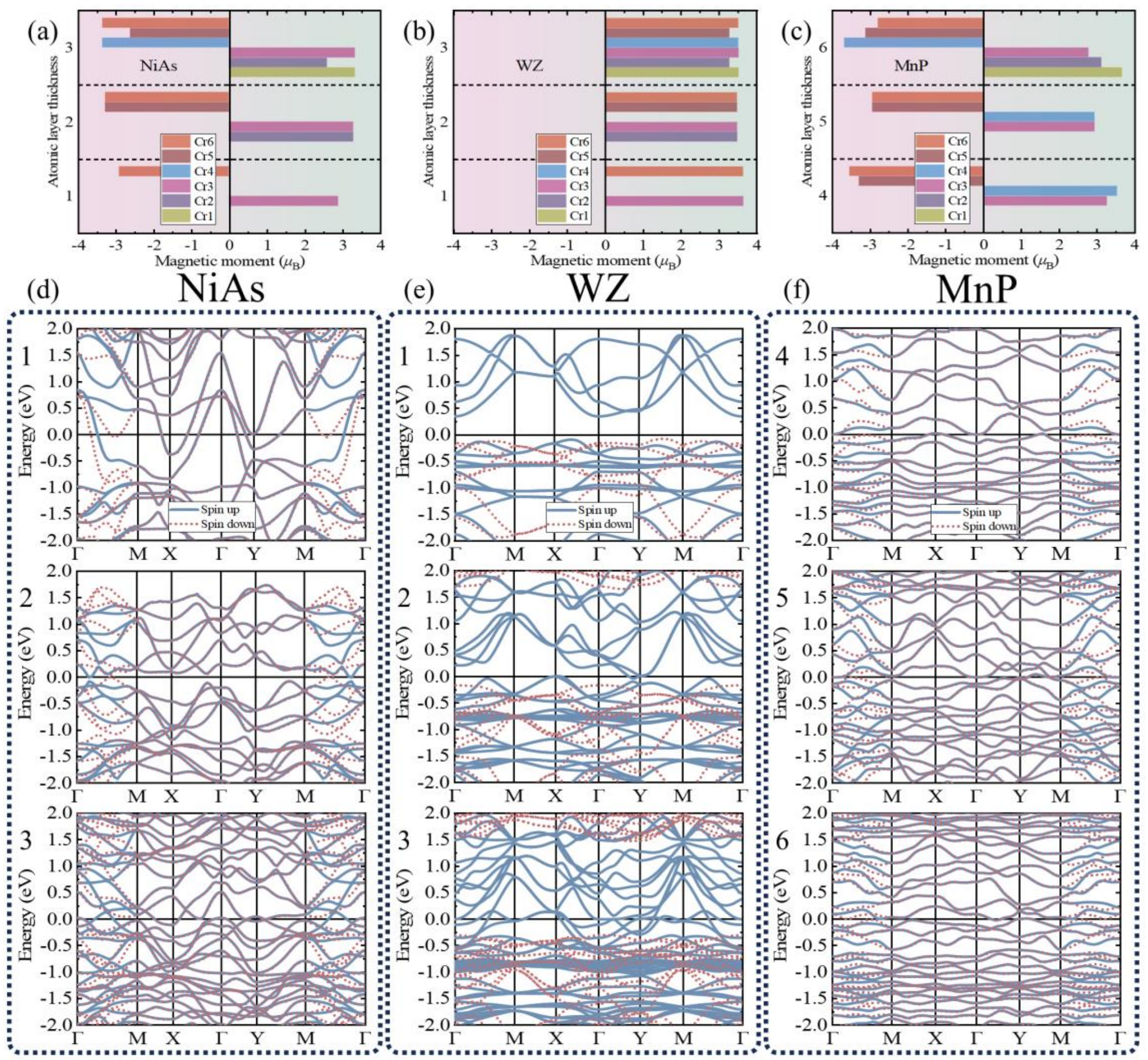

**Figure 9.** Atomic magnetic moments of Cr in CrSb (110) facet with varying atomic-layer thicknesses in the NiAs (a), WZ (b), and MnP (c) phases, and the corresponding spin-resolved band structures.

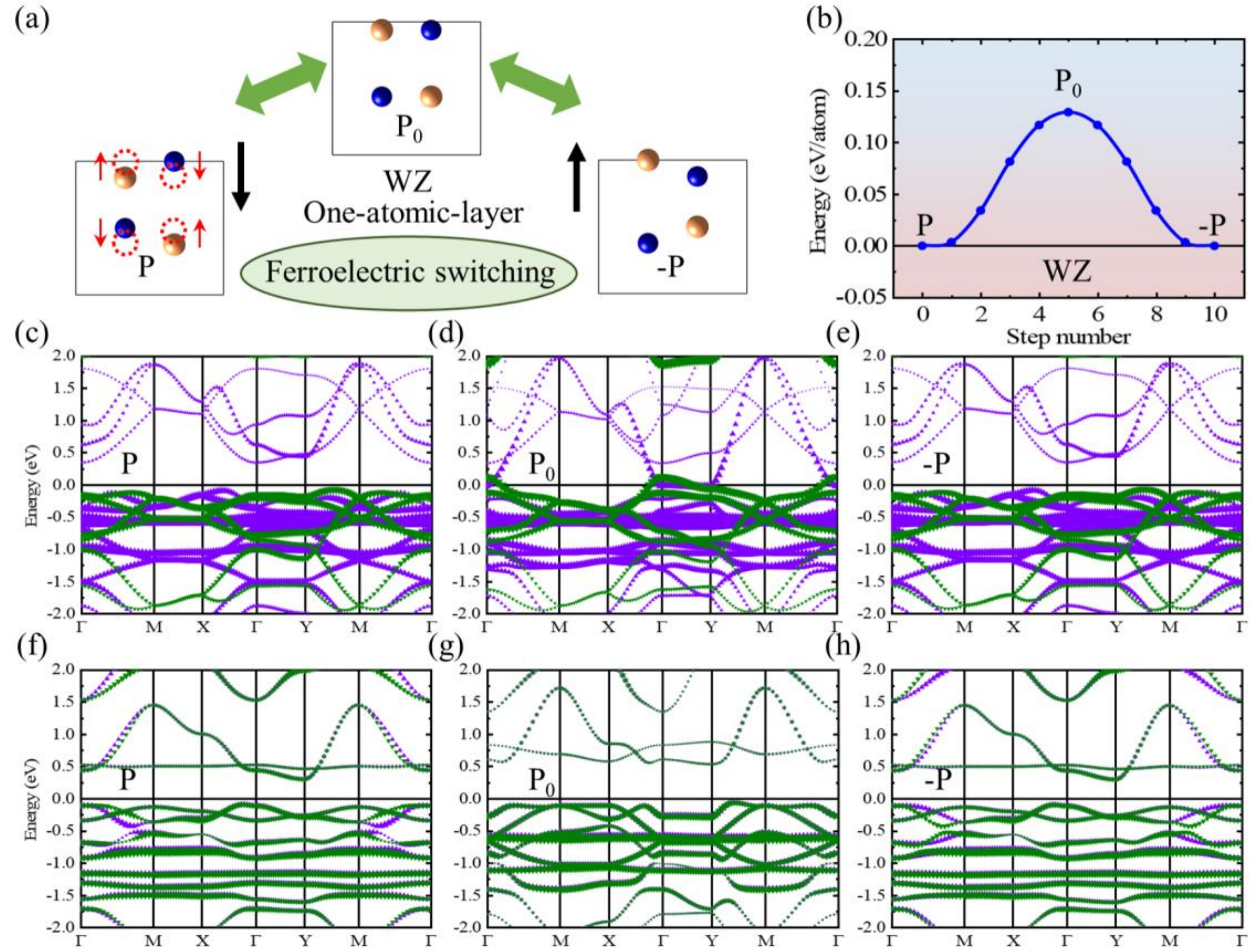

**Figure 10.** FE switching with atomic displacement paths (a) and energy barrier for the FE transition (b) for the WZ-phase CrSb (110) facet with one atomic layer. Spin-resolved and Cr-contributed band structures for the FM (c-e) and AFM (f-h) configurations in the polar (P, -P) and non-polar ($P_0$) states.

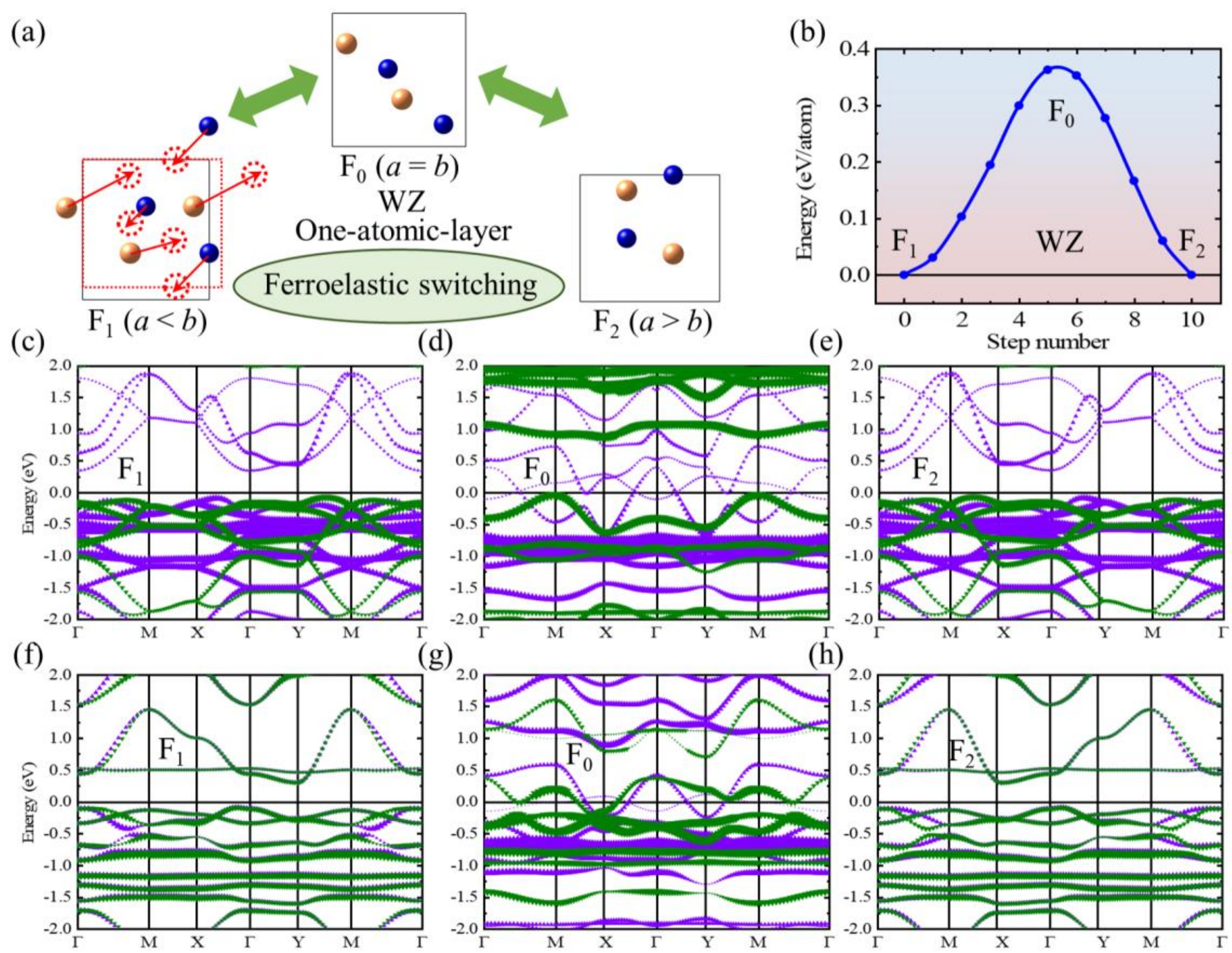

**Figure 11.** Ferroelastic (FC) switching with atomic displacement paths (a) and energy barrier for the FC transition (b) in the WZ-phase CrSb (110) facet with one atomic layer. Spin-resolved and Cr-contributed band structures for the FM (c-e) and AFM (f-h) configurations in the FC ($F_1$, $F_2$) and intermediate ($F_0$) states.

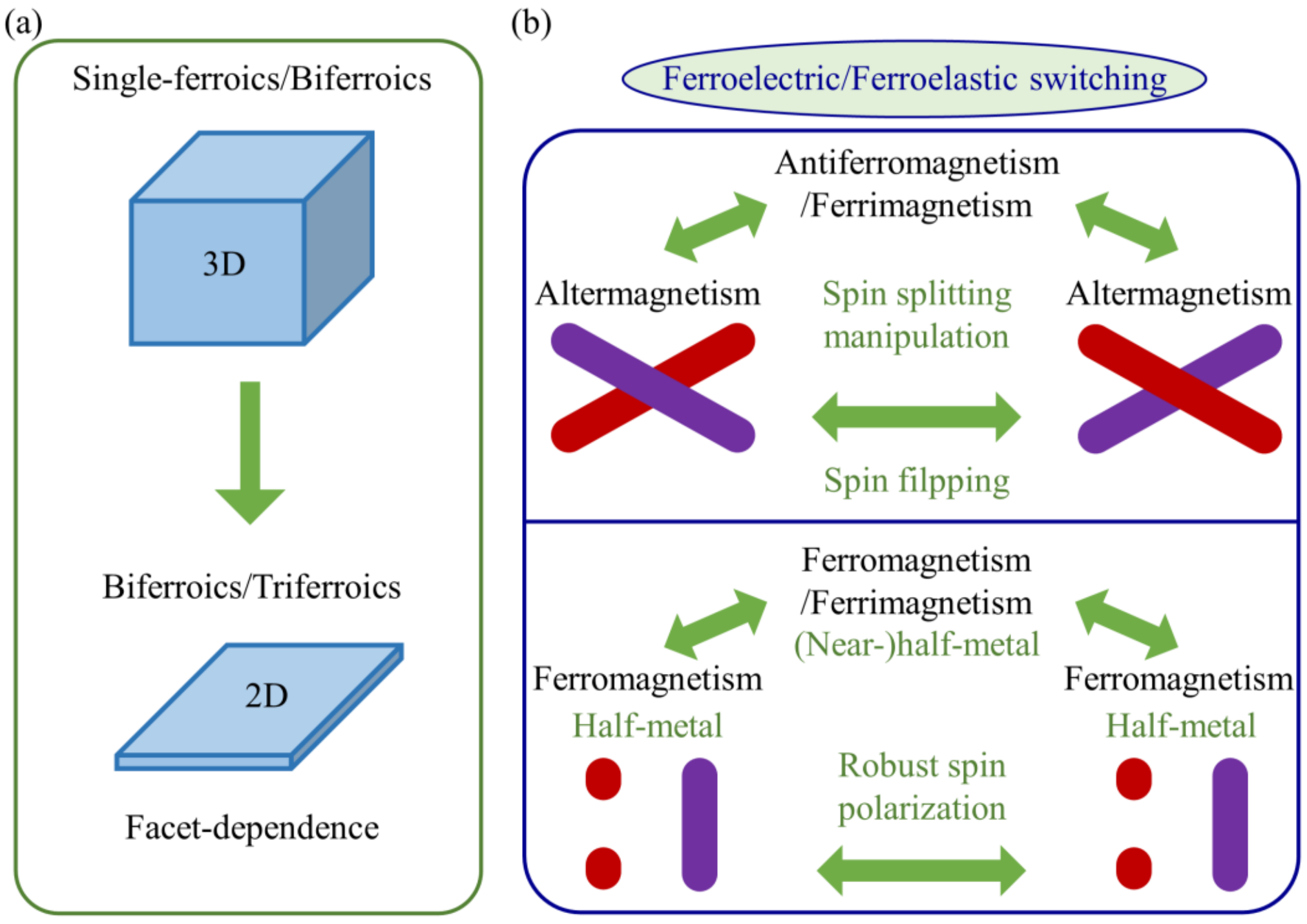

**Figure 12.** Schematic illustrations of the dimension- and facet-dependent biferroics and triferroics (a), and the manipulation of spin splitting via FE/FC switching in triferroic/biferroic CrSb (b).